\documentclass[a4paper,12pt]{article}
 \usepackage[dvips]{graphicx}
\begin{document}
 \begin{center}
 \Large\bf
 Rapidity equilibration and longitudinal expansion at RHIC\\[2.1cm]
 \large
 Georg Wolschin\footnote{E-mail: wolschin@uni-hd.de\hspace{5cm}
 http://wolschin.uni-hd.de}\\[.8cm]
 \normalsize\sc\rm
 Institut f\"ur theoretische Physik der Universit\"at,
 Philosophenweg 16,\\ D-69120 Heidelberg, Germany\\[2.6cm]
 %(Submitted Feb. 12, 2005)\\[2.0cm]
 \end{center}
 \bf
 \rm
 \bf{Abstract:}
 \normalsize\sc\rm
 The evolution of net-proton rapidity spectra with $\sqrt{s_{NN}}$ in
 heavy relativistic systems is proposed as an indicator
 for local equilibration and longitudinal expansion.
 In a Relativistic Diffusion Model, bell-shaped
 distributions in central collisions at AGS energies
 and double-humped nonequilibrium spectra at SPS show pronounced
 longitudinal collective expansion when compared to the available data.
 The broad midrapidity valley recently discovered at RHIC
 in central Au + Au collisions at $\sqrt{s_{NN}}$= 200 GeV indicates
 rapid local equilibration which is most likely due to deconfinement,
 and fast longitudinal expansion of the locally equilibrated subsystem.
 A prediction is made for Au + Au at $\sqrt{s_{NN}}$= 62.4 GeV.
 \newpage
 Net-baryon rapidity distributions have
 proven to be sensitive indicators for local equilibration and
 deconfinement in relativistic heavy-ion collisions \cite{wol03,wol04}.
 Whereas statistical model analyses of multiplicity ratios of produced
 particles \cite{mag01} appear to be consistent with the assumption that
 the system reaches equilibrium and can therefore be described by a
 temperature and a chemical potential, it is clear that one has to
 consider the distribution functions of the relevant observables in
 order to determine whether the system has indeed reached, or gone  
 through,
 thermal equilibrium.

 Distribution functions of produced particles in transverse momentum
 or energy have been shown in many analyses to be close to thermal
 equilibrium, if one takes into account the collective transverse
 expansion of the system \cite{bla99}. Here, the expansion
 velocities rise as the energy increases from AGS via SPS to RHIC when
 they are estimated based on thermal distribution functions.

 In longitudinal direction, however, it is more difficult to determine
 the degree of equilibration, and the collective expansion velocity. It
 is the main purpose of this note to provide a schematic macroscopic
 approach that allows to obtain both quantities. There have been
 earlier attempts to extract longitudinal collective velocities at AGS
 and SPS energies by assuming that the rapidity distributions
 are thermal, and that the remaining large difference to the data
 is then due to expansion \cite{sta96}.
 However, the net-baryon rapidity distributions at
 AGS, SPS and RHIC energies are clearly nonequilibrium distributions
 even in central collisions \cite{wol04}, and also the longitudinal
 distributions of produced particles are not fully thermalized
 \cite{wbs05}.

 To account for the nonequilibrium behavior of the system,
 the evolution of net-proton rapidity spectra with
 incident energy $\sqrt{s_{NN}}$= 4.9 to 200 GeV is studied
 analytically in a Relativistic Diffusion Model (RDM) \cite{wol04}, and
 compared to AGS \cite{ahl99}, SPS \cite{app99} and RHIC \cite{bea04}
 data. In addition to the nonequilibrium-statistical evolution
 as described in the RDM, collective longitudinal expansion
 is considered.

 The model \cite{wol04,lav02,ryb03,biy02} is based on a generalized
 Fokker-Planck equation (FPE)
 for the distribution function R(y,t) in rapidity space,
 $y=0.5\cdot ln((E+p)/(E-p))$,

 \begin{equation}
 \frac{\partial}{\partial t}[ R(y,t)]=-\frac{\partial}
 {\partial y}\Bigl[J(y)[R(y,t)]\Bigr]+D_{y}(t)
 \frac{\partial^2}{\partial y^2}[R(y,t)]^{2-q} .
 \label{fpe}
 \end{equation}
 with the nonextensivity-parameter q \cite{tsa88}, and the rapidity
 diffusion coefficient D$_{y}$ that contains the microscopic physics,
 and accounts for the broadening of the distribution functions through
 interactions and particle creations.

 The FPE can be solved analytically in the linear model case q=1,
 with constant D$_{y}$, a linear drift function
 \begin{equation}
 J(y)=(y_{eq}- y)/\tau_{y},
 \label{dri}
 \end{equation}
 and the rapidity relaxation time $\tau_{y}$. This is
 the so-called Uhlenbeck-Ornstein process for the relativistic
 invariant rapidity \cite{wol04,biy02}. The equilibrium value is $y_ 
 {eq}=0$
 in the center-of-mass for symmetric systems, whereas
 $y_{eq}$ is calculated from the given masses and
 momenta for asymmetric systems.

 Using $\delta-$function
 initial conditions at the beam rapidities for net baryons, analytical
 solutions of Eq.(\ref{fpe}) are obtained for various values
 of  $\tau_{{int}}/\tau_{y}$ (Fig.1 bottom as an example for Pb + Pb
 at SPS).
 Here the interaction time $\tau_{int}$
 is the time from nuclear contact to freeze-out, or the integration time
 of Eq.(\ref{fpe}). It determines how close to equilibrium the
 system can come, and it is obtained from dynamical models or
 from parametrizations of two-particle correlation measurements. In
 central Au + Au collisions at 200 A GeV, this yields
 $\tau_{int}\simeq 10 fm/c$ \cite{lis05}. For known interaction times,
 the rapidity relaxation times and diffusion coefficients
 can then be obtained from the data. Otherwise, $\tau_{{int}}/\tau_{y}$
 and the rapidity width coefficient
 \begin{equation}
 \Gamma_{y}=[8\cdot ln(2) \cdot D_{y}\cdot \tau_{y}]^{1/2}
 \label{gam}
 \end{equation}
 can be determined from the peak positions and width of the data,
 respectively, through the use of the equations for the mean values
 and the variances \cite{wol04} that are derived from Eq.(\ref{fpe}).

 The transport coefficients D$_{y}$ and $\tau_{y}$ are, however,
 macroscopically related to each other through a
 dissipation-fluctuation theorem with the equilibrium temperature T.
 This relation has been derived in \cite{wols99} from the requirement
 that the stationary solution of Eq.(\ref{fpe}) is equated
 with a Gaussian approximation to the thermal equilibrium
 distribution in y-space. At fixed incident energy, this weak-coupling
 rapidity diffusion coefficient turns out to be proportional
 \cite{wols99} to the equilibrium temperature T as
 in the analysis of Brownian motion (Einstein relation)
 \begin{equation}
 D_{y}\propto \frac{T}{\tau_{y}}.
 \label{ein}
 \end{equation}
 When compared to heavy-ion collision data at energies above the SIS
 region of 1-2 GeV, rapidity spectra that are calculated using
 this relation are consistently
 narrower \cite{wols99} than the data because collective processes
 (in particular, collective longitudinal expansion) are not included  
 in the
 weak-coupling dissipation-fluctuation theorem. Hence, the width
 coefficients $\Gamma_{y}$ as obtained from Eqs.(\ref{gam},\ref{ein}) are
 replaced by effective values $\Gamma_{y}^{eff}$. Depending on the
 system and the incident energy, they are typically factors of 2-5
 larger than the calculated values.

 In this work, I assume that the whole discrepancy
 $\Gamma_{y}^{eff}-\Gamma_{y}$
 is due to longitudinal expansion. I obtain
 the mean collective expansion velocity in $\pm z$-direction
 for baryons of rest mass  m$_{0}$ and relativistic mass m
 from the relativistic velocity
 \begin{equation}
 v=\sqrt{1-(m_{0}/m)^{2}}.
 \label{vel}
 \end{equation}
 For f degrees of freedom (f=1
 in rapidity space) the effective mass is
 \begin{equation}
 m=m_{0}+\frac{f}{2}\cdot(T_{eff}-T).
 \label{mass}
 \end{equation}
 In this expression, the mean energy content of the expansion is  
 written as
 $E_{coll}=f/2\cdot (T_{eff}-T)$ with an effective temperature $T_{eff}$
 and the equilibrium temperature T in analogy to the classical
 expression for the mean energy. Hence, the expansion enhances the
 particle mass $m_{0}$ to its relativistic value $m$.

 The thermal equilibrium temperature T should be associated
 with kinetic freeze-out. Typical values of T at RHIC energies
 are 110 MeV. This is significantly below the chemical freeze-out  
 temperature
 of about 170 MeV \cite{mag01} that is obtained from fits of hadron
 abundancies. In the weak-coupling case without
 collective effects, T$_{eff}$=T such that $v_{coll}=0$, and the
 distribution functions remain too narrow as compared to the data.
 With expansion, the diffusion coefficient is enlarged to its
 effective value, and since it is proportional to the temperature,
 $T_{eff}$ is obtained from $T$ using the same enhancement factor.

 With Eqs.(\ref{vel},\ref{mass}), the collective velocity becomes
 \begin{equation}
 v_{coll}^{||}=\left[1-\left[\frac{m_{0}}{m_{0}+\frac{f}{2}\cdot(T_ 
 {eff}-T)}
 \right]^{2}\right]^{1/2}
 \label{vco}
 \end{equation}
 with the limiting cases $v_{coll}^{||}=1$ for $T_{eff}>>T$, and
 $v_{coll}^{||}=0$ for $T_{eff}=T$.

 Comparing the solutions of Eq.(\ref{fpe}) to Au + Au central  
 collision (5
 per cent of the cross section) data at AGS-energies $\sqrt{s_{NN}}$=
 4.9 GeV \cite{ahl99}, it turns out that due to the relatively long
 interaction time $\tau_{int}$ and hence, the large ratio
 $\tau_{int}/\tau_{{y}}\simeq 1.08$ (Table I), the system is in
 rapidity space very close to thermal equilibrium, with longitudinal
 collective expansion at $v_{coll}^{||}$=0.49 (Fig.1, upper frame).
 The bell-shaped experimental distribution is in good agreement with
 the solution of Eq.\ref{fpe}. The distributions remain bell-shaped
 also at lower energies \cite{ahl99}.

 This situation changes at the higher SPS energy of $\sqrt{s_{NN}}$=
 17.3 GeV. Here net-proton Pb + Pb rapidity spectra corrected for hyperon
 feeddown \cite{app99} show two pronounced peaks in central collisions,
 which arise from the penetration of the incident baryons through the
 system. The gradual slow-down and broadening is described using Eq. 
 \ref{fpe}
 as a hadronic diffusion process with subsequent collective expansion,
 $v_{coll}^{||}$=0.75 (Table I). The associated nonequilibrium
 solutions with expansion for various values of $\tau_{int}/\tau_{{y}}$
 are shown in the lower frame of Fig.1. The system clearly does $not$  
 reach the
 dash-dotted equilibrium solution. Hence, both nonequilibrium
 properties, $and$ collective expansion are required to interprete the
 broad rapidity spectra seen at the SPS.

 Within the current framework, no indication for deconfinement of the
 incident baryons or other unusual processes can be deduced from the
 net-proton rapidity data at AGS and SPS energies, because the crucial
 midrapidity region is here too small.

 This is, however, different at RHIC energies $\sqrt{s_{NN}}$= 200
 GeV. The RDM nonequilibrium solution exhibits pronounced
 penetration peaks with
 collective longitudinal expansion $v_{coll}^{||}$=0.93, but it fails
 to reproduce the BRAHMS net-proton data \cite{bea04} in the broad
 midrapidity valley (solid curve in Fig.2, bottom): The diffusion of
 the incident baryons due to soft scatterings is not strong enough to
 explain the net baryon density in the central rapidity region.
 The individual nonequilibrium solutions $R_{1}$ and $R_{2}$ are  
 Gaussians, and
 if they fit the data points near y=±2, they necessarily grossly
 underpredict the midrapidity yield because it is in their tails.

 This central region can only be reached if a fraction of the system  
 undergoes a
 fast transition to local thermal equilibrium, dashed curve in Fig.2,
 bottom. With collective expansion of this locally equilibrated
 subsystem of 22 net protons ($v_{coll}^{||}$=0.93), the flat
 midrapidity BRAHMS data are well reproduced in an incoherent
 superposition of nonequilibrium and equilibrium solutions of Eq.\ref 
 {fpe}
 \begin{equation}
 \frac{dN(y,t=\tau_{int})}{dy}=N_{1}R_{1}(y,\tau_{int})+N_{2}R_{2}(y, 
 \tau_{int})
 +N_{eq}R_{eq}^{loc}(y).
 \label{normloc}
 \end{equation}
 The fast transition of a subsystem of $N_{eq}\simeq$55 baryons, or 22
 protons to local thermal equilibrium suggests that the associated  
 participant
 partons have reached local equilibrium through
 the sudden enhancement in the number of degrees of freedom that
 accompanies deconfinement. Microscopically, the large gap to
 midrapidity is thus bridged through hard scatterings of partons with
 subsequent thermal equilibration, rather than diffusion of nucleons.
 The fact that even hard partons can
 participate significantly in equilibration processes is evidenced
 by the high-$p_{T}$ suppression found in Au + Au at RHIC.

 The local equilibrium distribution $R_{eq}^{loc}$
 with expansion is a solution
 of Eq.(\ref{fpe}) for time to infinity with an enlarged (effective)
 diffusion coefficient
 that is related to the effective  equilibrium temperature.
 The nonequilibrium solutions $R_{1}, R_{2}$ of Eq.(\ref{fpe})
 are calculated with
 the same value of the diffusion coefficient. The underlying
 picture is that expansion affects nonequilibrium and
 equilibrium distributions in a similar fashion.

 In a schematic calculation for the lower RHIC energy of
 $\sqrt{s_{NN}}$= 62.4 GeV, I have used the rapidity diffusion
 coefficient from 200 GeV to obtain the result in the upper frame of  
 Fig.2.
 Two pronounced penetration peaks can be seen, together with a narrow
 midrapidity valley. The corresponding data have been taken, and are
 presently being analyzed by the BRAHMS collaboration \cite{oue04}.
 Once they are available, an adjustment of $\Gamma_{y}^{eff}$ and N$_{eq}^{loc}$  
 may be
 required. One may expect another correction because the total net
 proton number could change, although the baryon number is conserved.
 Note, however, that this change was found to be negligeable at SPS
 energies.

 The RDM-results can be further refined if the linear drift function
 is replaced by
 \begin{equation}
 J(y)=-\alpha\cdot m_{\perp}sinh(y)
 \label{drinl}
 \end{equation}
 with the transverse mass $m_{\perp}=\sqrt{m^2 + p_{\perp}^2}$, because
 this yields $exactly$ the Boltzmann distribution as the stationary
 solution of Eq.\ref{fpe} for q=1. This entails, however,
 to solve Eq.\ref{fpe}
 numerically. A numerical solution is also required if one tries to
 account for multiparticle effects through nonextensive statistics
 \cite{tsa88} with an explicit nonlinearity in the FPE, $1<q<1.5$. In
 this case, an approximate result may be obtained from a linear
 superposition of power-law solutions of Eq.\ref{fpe}
 \cite{lav02,ryb03,wol03}. The local equilibrium fraction is
 \begin{equation}
 R_{equ}^{loc,q}\propto [1-(1-q)\frac{m_{\perp}}{T}
 cosh(y)]^{\frac{1}{1-q}}
 \label{nesol}
 \end{equation}
 with
 $\int \limits_{-\infty}^{+\infty}R_{eq}^{loc,q}dy$ = 1. For
 $\sqrt{s_{NN}}$= 200 GeV, $m_{\perp}$= 1.1 GeV, T = 110 MeV,
 q=1.4 and N$_{eq}$=22
 protons, Eq.\ref{nesol} yields the dotted curve in Fig.2 (bottom).
 It lies in between the Boltzmann-Gibbs local equilibrium results
 without and with expansion. Hence, it simulates to some extent the
 collective effects.

 Whereas local equilibration in rapidity space occurs only
 for a small fraction of the
 participant baryons in central collisions at RHIC energies, the
 opposite is true for produced hadrons. This can be
 inferred from recent applications of the RDM with three sources
 (located at the beam rapidities, and at midrapidity) and
 $\delta$-function initial conditions \cite{wbs05}. About
 78\% of the produced hadrons in central Au + Au collisions at
 200 A GeV are
 found to be locally equilibrated in pseudorapidity space, whereas the
 corresponding number is 17\% in central d + Au.

 To conclude, I have interpreted recent results for
 central collisions of heavy systems at AGS, SPS and RHIC energies in
 a Relativistic Diffusion Model (RDM) for
 multiparticle interactions based on the interplay of
 nonequilibrium and local equilibrium ("thermal") solutions.
 In the linear version of the model, analytical results for the rapidity
 distribution of net protons in central collisions have been
 obtained and compared to data. The enhancement of the diffusion in  
 rapidity
 space as opposed to the expectation from the
 weak-coupling dissipation-fluctuation theorem has been interpreted as
 collective expansion, and longitudinal expansion velocities have been
 determined from a comparison between RDM-results and data based on a
 relativistic expression for the collective velocity. A prediction for  
 net-proton
 rapidity distributions at $\sqrt{s_{NN}}$= 62.4 GeV yields
 a smaller midrapidity valley than at 200 GeV
 which will soon be compared with forthcoming data.
 
 \newpage
 \rm
 TABLE I. Parameters for heavy relativistic systems at AGS, SPS and  
 RHIC-energies.
 The beam rapidity is expressed in the c.m. system.
 The fit parameter $\tau_{int}/\tau_{{y}}$ determines how fast the
 net-baryon system
 equilibrates in rapidity space. The second free parameter is the
 effective rapidity width coefficient $\Gamma_{y}^{eff}$. It
 includes the effect of expansion. The deduced
 longitudinal expansion velocity is $v_{coll}^{||}$.
 At 62.4 GeV, $\Gamma_{y}^{eff}$ will need adjustment(*) to forthcoming
 data.\\[1.5cm]
 \begin{tabular}{lccccccc}
 \hline\\
 Lab&$System$ & $\sqrt{s_{NN}}$ & $y_{b}^{c.m.}$ & $\tau_{{int}}/\tau_ 
 {y}$ &
 $ \Gamma_{y}^{eff}$
 &$v_{coll}^{||}$\\
 && (GeV) && & \\\\
 \hline\\
 AGS&Au + Au &   4.9 & 1.60 & 1.08 &  1.45 &0.49\cr\\
 SPS&Pb + Pb &  17.3 & 2.91 & 0.81 &  2.43 &0.75\cr\\
 RHIC&Au + Au &  62.4 & 4.20 & 0.34 &  2.94* &0.86\cr\\
 RHIC&Au + Au & 200 & 5.36 & 0.26 & 4.29& 0.93\cr\\
 \hline
 \end{tabular}
 \newpage
 \Large\bf
 Figure captions
 \normalsize\rm
 \begin{description}
 \item[FIG. 1:]
 Net-proton rapidity spectra in the Relativistic Diffusion Model
 (RDM), solid curves. Central Au + Au at AGS energies $\sqrt{s_{NN}}$ =
 4.9 GeV (data from \cite{ahl99}) is close to a longitudinally expanding
 equilibrium distribution (top). Central Pb + Pb at SPS energies $\sqrt 
 {s_{NN}}$ =
 17.3 GeV (data from \cite{app99}) remains relatively far from thermal
 equilibrium and therefore, double-humped with two penetration peaks
 (middle). Dashed curves are thermal equilibrium distributions without
 collective expansion, dash-dotted curves with expansion.
 The bottom panel shows analytical Pb + Pb RDM-solutions
 for $\tau_{{int}}/\tau_{y}$= 0.08, 0.1, 0.2, 0.5,
 0.8, 1.0, 1.1, 1.3, all with longitudinal collective expansion, Table I.
 The transition from bell-shaped (AGS) to
 double-humped (SPS) is clearly shown in the RDM.
 \item[FIG. 2:]
 Net-proton rapidity spectra in the RDM (solid curves) for central  
 collisions of
 Au + Au at RHIC energies $\sqrt{s_{NN}}$ = 62.4 GeV (top), and 200  
 GeV (middle,
 central 5\% of the cross section)
 compared to BRAHMS data \cite{bea04} at the higher energy. The net
 proton content is 158. The high midrapidity yield is not attainable
 through soft scatterings in the Relativistic Diffusion Model. Instead,
 this region containing $\simeq22$ protons at 200 GeV is populated  
 through hard
 scatterings with ensuing local equilibration (dashed curve), and
 collective expansion (dashed areas). See Table I for parameters.
 The dotted curve is the nonlinear local equilibrium solution
 of the FPE for q=1.4, cf. text.
 \end{description}
 \newpage
 \vspace{1cm}
 \includegraphics[width=10cm]{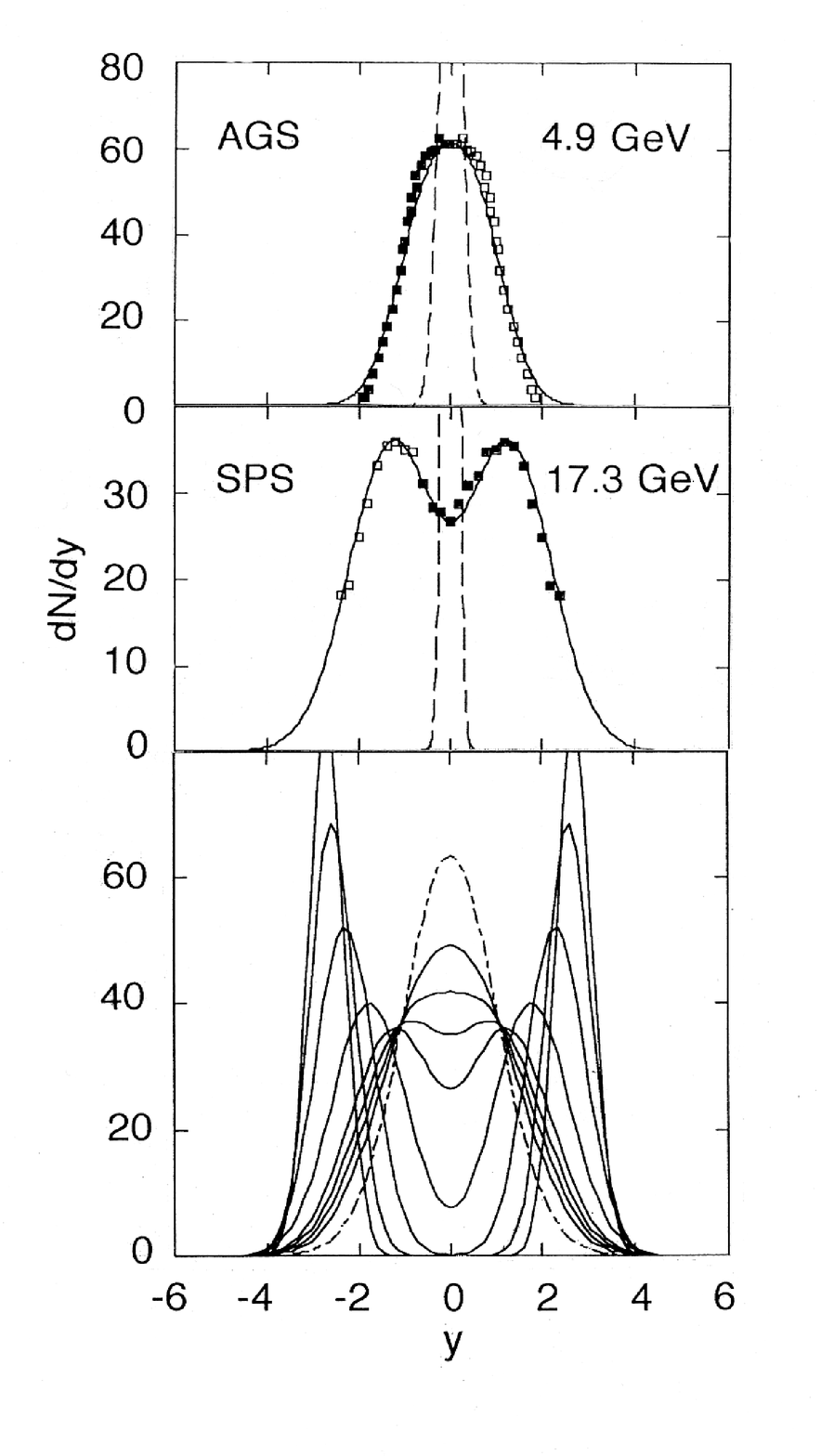}
 \newpage
 \vspace{1cm}
 \includegraphics[width=10cm]{fig2.eps}
 \end{document}